\documentclass[12pt,preprintnumbers,aps,amssymb,nofootinbib,amsmath]{revtex4}
\usepackage{epsfig,epsf}
\usepackage{graphicx}
\usepackage{epstopdf}
\newcommand{\beq}{\begin{equation}}
\newcommand{\eeq}{\end{equation}}
\def\eq#1{{(\ref{#1})}}

\def\b#1{\mathbf{#1}}
\newcommand{\im}{\mathrm{Im}}

\newcommand{\e}{\varepsilon}

\newcommand{\bea}{\begin{eqnarray}}
\newcommand{\ena}{\end{eqnarray}}
\newcommand{\ben}{\begin{eqnarray*}}
\newcommand{\een}{\end{eqnarray*}}

\begin{document} 

\preprint{BNL-NT-07/47}
\preprint{RBRC-703}

\title{Universal properties of bulk viscosity\\ near the QCD phase transition}

\author{Frithjof Karsch$^a$, Dmitri Kharzeev$^a$ and Kirill Tuchin$^{b,c}$}
 
\affiliation{
a) Department of Physics, Brookhaven National Laboratory,\\
Upton, New York 11973-5000, USA\\
b) Department of Physics and Astronomy,\\
Iowa State University, Ames, Iowa, 50011, USA\\
c) RIKEN BNL Research Center,\\
Upton, New York 11973-5000, USA\\
}

\date{\today}
\pacs{}

\begin{abstract} 
We extract the bulk viscosity of hot quark-gluon matter in the presence of light quarks from the recent lattice data on the QCD equation of state. For that purpose we extend the 
sum rule analysis by including the contribution of light quarks. We also discuss the universal properties
of bulk viscosity in the vicinity of a second order phase transition, as it might occur in the chiral limit of QCD at fixed strange quark mass and most likely does occur in two-flavor QCD. 
We point out that a chiral transition in the  $O(4)$ universality class at zero baryon density as well as the transition at the chiral critical point which belongs to the $Z(2)$ universality class both lead to the critical behavior of bulk viscosity. In particular, the latter universality class implies the divergence of the bulk viscosity, which may be used as a signature of the critical point.
We discuss the physical picture behind the dramatic increase of bulk viscosity seen in our analysis, and devise possible experimental tests of related phenomena.

\end{abstract}

\maketitle

\section{Introduction}

Recently, it was pointed out that bulk viscosity in $SU(3)$ gluodynamics rises dramatically 
in the vicinity of the deconfinement phase transition \cite{Kharzeev:2007wb}. The analysis was based on an exact sum rule derived from low--energy theorems of broken scale invariance, the high-statistics lattice data on the equation of state \cite{Boyd:1996bx}, and an ansatz for the spectral density of the correlation function of the energy-momentum tensor. The rapid increase of bulk viscosity has been associated with the fast growth of the thermal expectation value 
of the trace of the energy-momentum tensor $\langle \theta \rangle_T = \mathcal{E} - 3 P$, where $\mathcal{E}$ is the energy density and $P$ is the pressure. 
Very recently, bulk viscosity in $SU(3)$ gluodynamics has been measured on the lattice by the direct analysis of the correlation function of the trace of the energy-momentum tensor \cite{Meyer:2007dy} (see also \cite{Nakamura:2007cm}). 
The numerical results of the study \cite{Meyer:2007dy} agree very well with the previous analysis \cite{Kharzeev:2007wb}. Since the methods used in Refs.~\cite{Kharzeev:2007wb,Meyer:2007dy}, and the lattice observables used as an input, are quite different, this agreement indicates that the dramatic growth (by about three orders of magnitude!) of bulk viscosity in the vicinity of the deconfinement phase transition is indeed a prominent feature of $SU(3)$ gluodynamics.    

Is this behavior specific for the first order phase transition in $SU(3)$ gluodynamics?
What is the ``microscopic" dynamics responsible for the growth of bulk viscosity? Can this growth 
be classified in terms of the general theory of critical phenomena? Can it be used in the experimental studies to isolate the signatures of phase transitions in heavy ion collisions?
In this paper we address these questions by generalizing the previous analysis to the case of 
QCD with two light quarks and a strange quark, which is of direct relevance to heavy ion experiments. 
We will use as an input the very recent high statistics lattice data on the equation of state of QCD with almost physical quark masses from the 
RIKEN-BNL-Columbia-Bielefeld Collaboration \cite{Cheng:2007jq}.

 The paper is organized as follows. In section \ref{kubosec} we review the definition of bulk viscosity and the formalism relating it to the correlation function of the trace of the energy-momentum tensor based on Kubo's formula. In section \ref{lowsec} we discuss the low-energy theorems based on QCD renormalization group which constrain the zero-momentum correlation functions involving the trace of the energy-momentum tensor. In this Section we generalize the sum rule of Ref.~\cite{Kharzeev:2007wb} to full QCD with $2+1$ flavors.  In Section \ref{latsec} we briefly review the relevant lattice thermodynamics results, introduce and motivate our ansatz for the spectral density, and use the sum rule to extract the bulk viscosity near the critical temperature. We find that the behavior of bulk viscosity in full QCD is qualitatively similar to the case of $SU(3)$ gluodynamics.  We then proceed to the discussion of uncertainties associated with our method, and estimate the error bars which should be associated with our result.
 In Section \ref{unisec} we classify the critical behavior responsible for the rapid growth of bulk viscosity 
 near the phase transition. We argue that at zero baryon density, this behavior belongs to the 
 $O(4)$ universality class. The bulk viscosity in this case does not diverge at $T=T_c$, but has a cusp.
 On the other hand, in the vicinity of the chiral critical point at finite baryon density, the critical behavior is in the $Z(2)$ universality class. This means that the bulk viscosity should {\it diverge} at $T=T_c$. 
 This behavior should manifest itself in heavy ion collisions through the decrease of average transverse momentum of produced particles, accompanied by the increase in total multiplicity. This is due to both the increase in entropy associated with a large bulk viscosity, and the associated quenching of the transverse hydrodynamical  expansion of the system (``radial flow"). Finally, we summarize our findings and discuss the qualitative physical picture which arises from our analysis. In particular, we argue  
 that the rapid growth of bulk viscosity favors the scenario of ``soft statistical hadronization" in which the expanding system hadronizes at the phase transition by producing a large number of color screening soft partons.  
 
\section{Kubo's formula for bulk viscosity}\label{kubosec}
According to Kubo's formula  the shear $\eta$ and the bulk $\zeta$ viscosities are related to the correlation function of the stress tensor $\theta_{ij}(x)$,  $i,j=1,2,3$, as 
\bea\label{def.visc}
&& \eta(\omega)\left( \delta_{il}\delta_{km}+\delta_{im}\delta_{kl}-\frac{2}{3}\delta_{ik}\delta_{lm}\right)
+\zeta(\omega)\delta_{ik}\delta_{lm}\nonumber\\
&&=\frac{1}{\omega}\lim_{\bf k\to 0}\int d^3x
\int_0^\infty dt\, e^{i(\omega t-\bf k\bf r)}\langle [\theta_{ik}(t,\b r),\theta_{lm}(0)]\rangle 
\ena
Contracting the $i,k$ and $l,m$ indices gives the bulk viscosity as a static limit of the correlation function of the trace of the stress  tensor 
\beq\label{kubo0}
\zeta = \frac{1}{9}\lim_{\omega\to 0}\frac{1}{\omega}\int_0^\infty dt \int d^3r\,e^{i\omega t}\,\langle [\theta_{ii}(x),\theta_{kk}(0)]\rangle \,.
\eeq
This formula can be written in terms of  Lorentz-invariant operators  if we recall that the ensemble average of the commutator of the Hamiltonian $H$   with any operator $\mathfrak{O}$ in \emph{equilibrium} is given by 
\beq
\left\langle \left[\int d^3x\,\theta_{00}(x),\mathfrak{O}\right]\right\rangle_\mathrm{eq}=\langle[H,\mathfrak{O}]\rangle_\mathrm{eq}=i\left\langle \frac{\partial \mathfrak{O}}{\partial t}\right\rangle _\mathrm{eq}=0\,.
\eeq
Thus we can write \eq{kubo0} as 
\beq\label{kubo}
\zeta = \frac{1}{9}\lim_{\omega\to 0}\frac{1}{\omega}\int_0^\infty dt \int d^3r\,e^{i\omega t}\,\langle [\theta^\mu_\mu(x),\theta^\mu_\mu(0)]\rangle \,.
\eeq

In the following  we are going to calculate the bulk viscosity using the low energy theorems involving the Green's functions of the trace of energy-momentum tensor $\theta(x)$. Therefore, it is convenient to re-write  Kubo's formula (\ref{kubo}) using  the retarded Green's function as
\beq\label{kubo2}
\zeta=\frac{1}{9}\lim_{\omega\to 0}\frac{1}{\omega}\int_0^\infty dt\int d^3r\,e^{i\omega t} \, iG^R(x)\,=\,\frac{1}{9}\lim_{\omega\to 0}\frac{1}{\omega} \,iG^R(\omega,\vec 0)\,=\,-\frac{1}{9}\lim_{\omega\to 0}\frac{1}{\omega} \,\im G^R(\omega,\vec 0)\,.
\eeq
The last equation follows from the fact that due to P-invariance, the function $\im G^R(\omega,\vec 0)$ is odd in $\omega$ while $\mathrm{Re}\,G^R(\omega,\vec 0)$ is even in $\omega$.  Note that the non-vanishing $\zeta$ implies the existence of a \emph{massless} excitation in the spectral density $\rho$. 

Let us define the spectral density
\beq\label{spec.den}
\rho(\omega,\vec p)=-\frac{1}{\pi}\,\im G^R(\omega,\vec p)\,.
\eeq
Using the Kramers-Kronig relation the retarded Green's function can be represented 
as
\beq\label{kk}
G^R(\omega,\vec p)=\frac{1}{\pi}\int_{-\infty}^\infty\,\frac{\im G^R(u,\vec p)}{u-\omega -i\e}\,du=
\int_{-\infty}^\infty\,\frac{\rho(u,\vec p)}{\omega -u+i\e}\,du
\eeq

The retarded  Green's function $G^R(\omega,\vec p)$ of a bosonic excitation is related to the Euclidean Green's function $G^E(\omega,\vec p)$ by analytic continuation
\beq\label{analit}
G^E(\omega,\vec p)=-G^R(i\omega,\vec p)\,,\quad \omega >0\,.
\eeq
Using \eq{kk} and the fact that $\rho(\omega,\vec p)=-\rho(-\omega,\vec p)$ we recover
\beq\label{ge00}
\mathfrak{G}\equiv \lim_{\omega\to 0}G^E(\omega,\vec 0)=2\int_0^\infty \frac{\rho(u,\vec 0)}{u}\,du\,.
\eeq
In the next section we use the low energy theorems to relate $\mathfrak{G}$ to the energy density and the pressure of  hot hadronic matter. 

\section{Low energy theorems at finite $T$}\label{lowsec}

In a conformally invariant theory the trace of the energy-momentum tensor  vanishes. Therefore,  Eq.~\eq{kubo} implies that the bulk viscosity is a measure for the violation of conformal invariance. QCD is a conformally invariant theory at the classical level. However, quantum fluctuations generate the scale anomaly which manifests itself in the non-conseravtion of the dilatational current $s_\mu$
 \cite{scale1,scale2}:
\beq\label{trace}
\partial^\mu s_\mu=\theta^\mu_{\;\mu}=m_q\bar qq+\frac{\beta(g)}{2g}
\,F^2\equiv \theta_F+\theta_G\,,
\eeq
where we adopted the shorthand notation $F^2\equiv F^{a\,\mu\nu}F_{\mu\nu}^a$ and $q,\bar q$ are three-vectors in the quark flavor space (i.e. summation over flavors is assumed). 
$\beta(g)$ is the QCD $\beta$-function, which governs the behavior of the running coupling
\beq
\mu \frac{d g(\mu)}{d \mu} = \beta (g)\,. \label{rg}
\eeq   
At the leading order $\beta = -bg^3/(4\pi)^2$.

Although the scale symmetry of the QCD lagrangian is broken by quantum vacuum fluctuations, there is a remaining symmetry imposed by the requirement of the renormalization group invariance of observable quantities \cite{Novikov:1981xj}. This symmetry manifests itself in a chain of low energy theorems (LET) for the correlation functions of the operator $\theta_G(x)=(\beta/2g)F^2(x)$.
 These low-energy theorems entirely determine the dynamics of the effective low-energy theory. This effective theory has an elegant geometrical interpretation \cite{Migdal:1982jp}; in particular, gluodynamics can be represented as a classical theory formulated on a curved (conformally flat) space-time background \cite{KLT-Pom}.   At finite temperature, the breaking of scale invariance by quantum fluctuations results in $\theta=\mathcal{E}-3P\neq 0$ clearly observed on the lattice for $SU(3)$ gluodynamics \cite{Boyd:1996bx}; the presence of quarks \cite{Bernard:1996cs} including the physical case of two light and a strange quark \cite{Cheng:2007jq,Bernard:2006nj,Aoki}, or  
considering large $N_c$ \cite{Bringoltz:2005rr} does not change this conclusion. 

The LET of  Ref.~\cite{Novikov:1981xj,Migdal:1982jp} were generalized to the case of finite temperature in \cite{Ellis:1998kj,Shushpanov:1998ce} (lattice formulation has been discussed very recently in Ref. \cite{Meyer:2007fc}). The lowest in the chain of relations reads (at zero baryon chemical potential):
\beq\label{let0}
\int d^4 x \,\langle T\theta_G(x),\mathfrak{O}(0)\rangle =\left(T\frac{\partial}{\partial T} -d\right)\,\langle \mathfrak{O}\rangle_T\,,
\eeq
where $d$ is a canonical dimension of the operator $\mathfrak{O}$. In particular, 
\begin{subequations}
\beq\label{let}
\int d^4x\,\langle T\theta_G(x),\theta_G(0)\rangle =\left(T\frac{\partial}{\partial T} -4\right)\,\langle \theta_G\rangle_T\,,
\eeq
\beq\label{let1}
\int d^4x\,\langle T\theta_G(x),\theta_F(0)\rangle =\left(T\frac{\partial}{\partial T} -3\right)\,\langle \theta_F\rangle_T\,,
\eeq
\end{subequations}
where $\theta_G$ and $\theta_F$ are the contributions to the trace of energy-momentum tensor from the gauge field and the quarks correspondingly defined in \eq{trace}.

Using \eq{trace} and \eq{let},\eq{let1}  we obtain
\bea\label{bra}
\int d^4x  \,\langle T\theta(x),\theta(0)\rangle&=&
\left(T\frac{\partial}{\partial T} -4\right)\,\langle \theta_G\rangle_T+2\left(T\frac{\partial}{\partial T} -3\right)\,\langle \theta_F\rangle_T+
\int d^4x \,\langle T\theta_F(x),\theta_F(0)\rangle\nonumber\\
&\approx& \left(T\frac{\partial}{\partial T} -4\right)\,\langle \theta\rangle_T+\left(T\frac{\partial}{\partial T} -2\right)\,\langle \theta_F\rangle_T\,,
\ena
where in the last line we neglected the  $\langle T\theta_F(x),\theta_F(0)\rangle$ correlation function which is proportional to $m^2$. We will see in the next section that the quark mass $m$ can indeed be treated as a small parameter \footnote{Note that the heavy quark terms decouple in the low-energy matrix elements of the energy-momentum tensor canceling against the heavy quark part of the beta-function in the $\theta_G$ operator \cite{Novikov:1981xj}.}. 

To relate the thermal expectation value of $\langle \theta_G\rangle_T$ to the quantity $(\mathcal{E}-3P)^*$ computed on the lattice, we should keep in mind that
\beq\label{renor}
(\mathcal{E}-3P)^*= \langle \theta\rangle_T - \langle \theta\rangle_{0},
\eeq
i.e. the zero-temperature expectation value of the trace of the energy-momentum tensor
\beq\label{zerotem}
\langle \theta\rangle_{0} = - 4 |\epsilon_v|
\eeq
has to be subtracted; it is related to the vacuum energy density $\epsilon_v < 0$.
Analogously, 
\beq\label{renor1}
\langle \theta_F\rangle_T=\langle m \bar q q\rangle^* + \langle m \bar q q\rangle_0\,.
\eeq

Since the lattice data we are going to use correspond to almost physical quark masses, we can use the PCAC relations to express the zero-temperature vacuum expectation value $\langle m \bar q q\rangle_0$ of the scalar quark operator through the pion and kaon masses $M_\pi, M_K$ and decay constants $f_\pi, f_K$:
\beq\label{pcac}
\langle m \bar q q\rangle_0  = - M_\pi^2\ f_\pi^2 - M_K^2\ f_K^2
\eeq
Eq.~\eq{ge00} implies that the l.h.s. of \eq{bra} is just the constant $\mathfrak{G}$. Using \eq{let},\eq{let1} and \eq{renor},\eq{renor1},\eq{pcac} in the r.h.s. of \eq{bra} we derive the following sum rule
\bea
&&2\int_0^{\infty} \frac{\rho(u,\vec 0)}{u}\,du =\left(T\frac{\partial}{\partial T} -4\right)\, \{ (\mathcal{E}-3P)^*+\langle\theta\rangle_0\}+
 \left(T\frac{\partial}{\partial T} -2\right)\, \{ \langle m \bar q q\rangle^*+\langle \theta_F\rangle_0 \} \nonumber\\
 &&= T\, s\, \left(\frac{1}{c_s^2}-3\right)\,-\,4(\mathcal{E}-3P)\,
 +\, \left(T\frac{\partial}{\partial T} -2\right)\,\langle m \bar q q\rangle^*
+16|\epsilon_v|  -6\langle m\bar q q\rangle_0 \label{sum} \\
&&= T\, s\, \left(\frac{1}{c_s^2}-3\right)-4(\mathcal{E}-3P)+\left(T\frac{\partial}{\partial T} -2\right)\langle m \bar q q\rangle^*
+16|\epsilon_v| 
+  6 (  M_\pi^2\ f_\pi^2 + M_K^2\ f_K^2)\, \label{sum1}
\ena
where $s$ is the entropy density and $c_s$ is the speed of sound.  
In deriving  \eq{sum} we used basic thermodynamic
relations to relate temperature derivatives of energy density and pressure
to the entropy density, $s=\partial P/\partial T$, specific heat,
$c_v=\partial {\mathcal E}/\partial T$ and the
velocity of sound, $c_s^2 = \partial P/\partial {\mathcal E} = s/c_v$. Eq.~\eq{sum1} is the main result of our paper.

\section{Bulk viscosity from the lattice data}\label{latsec}

\subsection{Spectral density}

In order to extract the bulk viscosity $\zeta$ from (\ref{sum1}) we need to make an ansatz for the spectral 
density $\rho$. At high frequency, the spectral density should be described by perturbation theory; however, 
the perturbative (divergent) contribution has been subtracted in the definition of the quantities on the r.h.s.\ 
of the sum rule (\ref{sum}), and so we should not include the high frequency perturbative continuum $\rho(u) \sim \alpha_s^2\, u^4$ on the l.h.s.\ as well.  Indeed, when the frequency $\omega$ is much larger than the temperature of the system $T$, $\omega \gg T$ the spectral density $\rho(\omega)$ should be temperature-independent. Therefore the subtraction of the expectation value of the trace of the energy-momentum tensor performed on the lattice should 
remove the high frequency perturbative part of the spectral density.

Note that the subtraction of the zero-temperature expectation value $\langle \theta\rangle_{0}$ \eq{zerotem} removes the ultra-violet (UV) singularity $1/a^4$ ($a$ is the lattice spacing) in \eq{renor}, so that the resulting quantity $(\mathcal{E}-3P)^*$ is UV finite. Such a subtraction implies the removal of local contact terms in the renormalized correlation function of the trace of the energy-momentum tensor.

In the small frequency region, we will {\it assume}  
 the following ansatz
\beq\label{ansatz}
\frac{\rho(\omega,\vec 0)}{\omega}=\frac{9\,\zeta}{\pi}\frac{\omega_0^2}{\omega_0^2+\omega^2}\,,
\eeq
which satisfies \eq{kubo2} and \eq{spec.den}. Substituting (\ref{ansatz}) in \eq{sum} we arrive at 
\beq\label{ze}
9\,\omega_0\,\zeta = T\, s\, \left(\frac{1}{c_s^2}-3\right)-4(\mathcal{E}-3P)+\left(T\frac{\partial}{\partial T} -2\right)\langle m \bar q q\rangle^*
+16|\epsilon_v| 
+  6 (  M_\pi^2\ f_\pi^2 + M_K^2\ f_K^2)
\eeq

The parameter $\omega_0 = \omega_0(T)$ is a scale at which the perturbation theory becomes valid. On dimensional grounds, we expect it to be proportional to the temperature, $\omega_0  \sim T$. We estimate it as the scale at which the lattice calculations of the running coupling \cite{Kaczmarek:2004gv} coincide with the perturbative expression at a given temperature (see discussion below). 

\begin{figure}[ht]
  \begin{tabular}{cc} 
      \includegraphics[width=8cm]{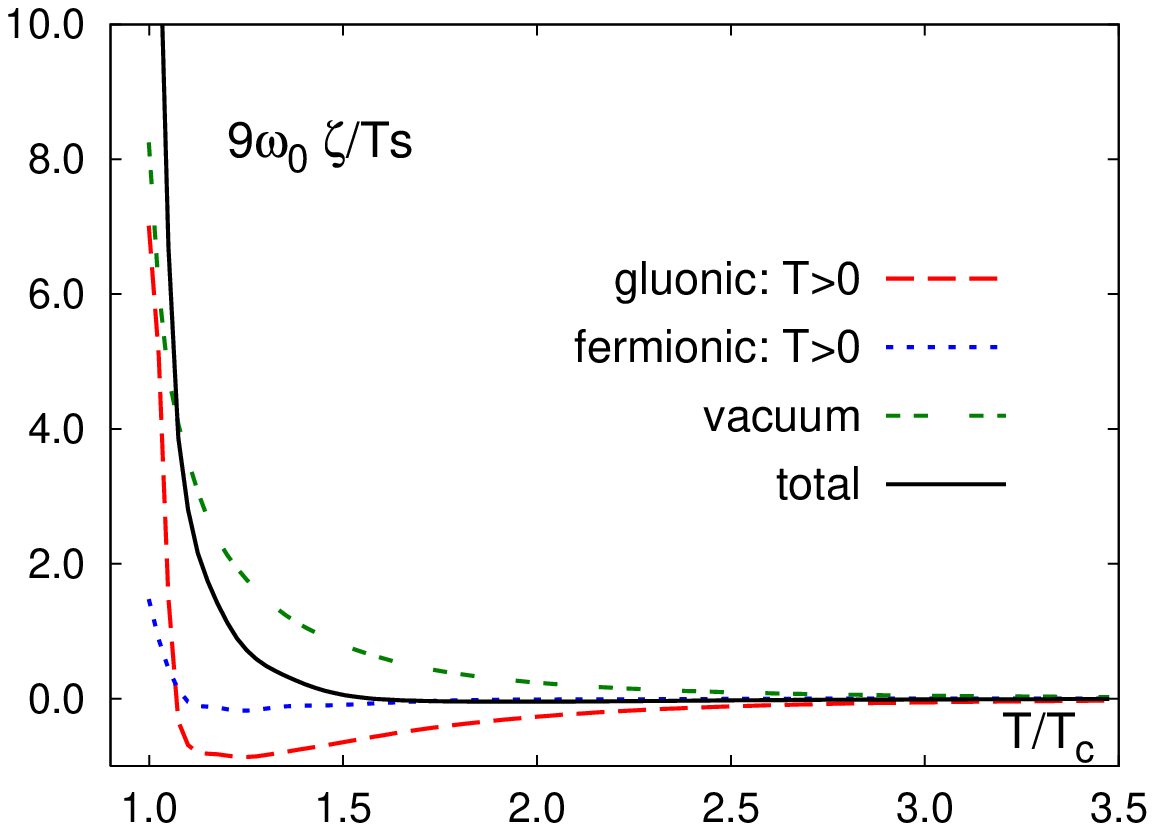} &
        \includegraphics[width=8cm]{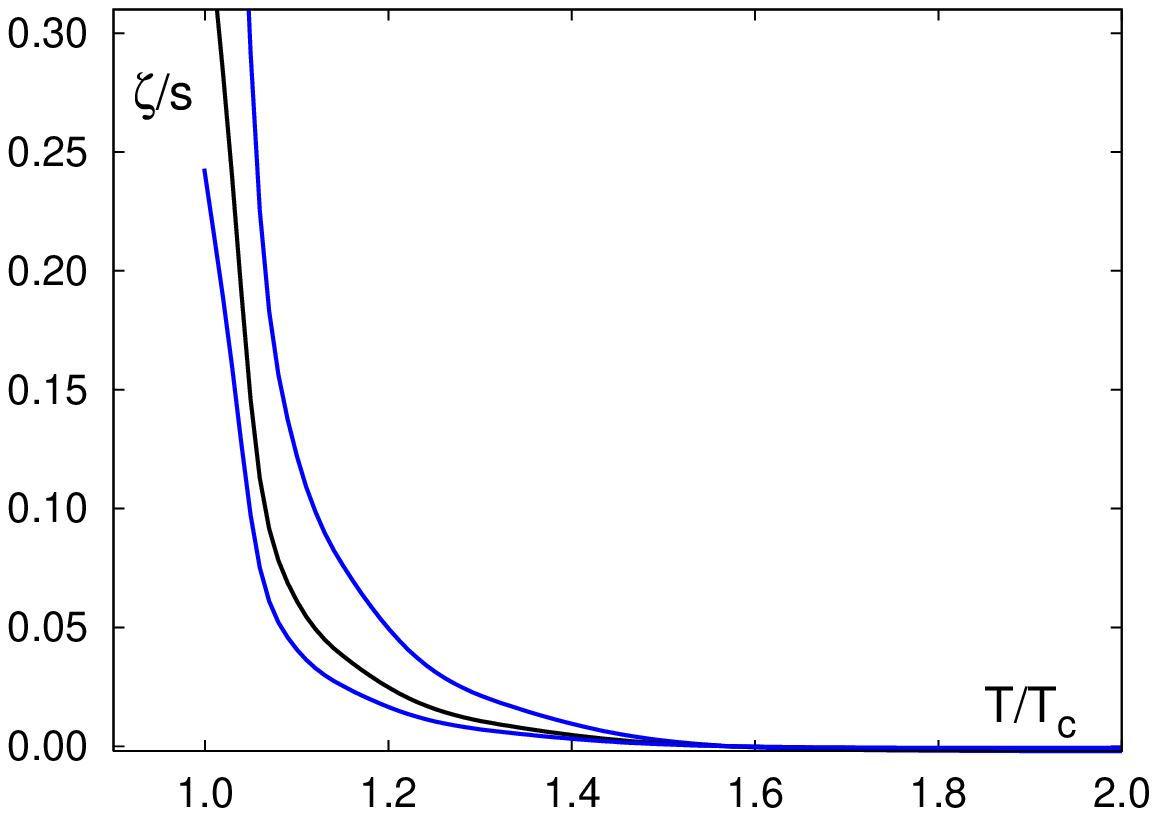}\\
    (a) & (b) 
\end{tabular}
\caption{Bulk viscosity in the high temperature phase of QCD versus temperature in
units of the transition temperature. Part (a) of the figure
shows various contributions to $9\omega_0(T)\zeta/Ts$ given in \eq{ze}.
In part (b) of the figure we show the bulk viscosity in units
of the entropy density for $\omega_0 = 0.5,\, 1,\, 1.5$~GeV (top to bottom) which reflects the uncertainty in the determination of this scale parameter. }
\label{fig:viscosity}
\end{figure}

We note that the expression for bulk viscosity  \eq{ze} consists of a
thermal part that can be determined through lattice calculations, and a
vacuum contribution, which we fixed using PCAC relations for the quark
condensates and using the gluon condensate value $|\epsilon_v|^{1/4} =  250\ {\rm MeV}$. 
With this the
vacuum part is given numerically by 
\begin{equation}
16 |\epsilon_v| (1 + \frac{3}{8} \cdot 1.6) \simeq (560\ {\rm MeV})^4
\simeq (3 \,T_c)^4 \, .
\label{vacuum}
\end{equation}
Using the lattice results for trace anomaly, $({\mathcal E} -3P)/T^4$, the fermionic contribution to it as well as the square of the velocity of sound presented in Ref. \cite{Cheng:2007jq} we can determine the bulk
viscosity. To be specific we use the data set obtained from simulations
on lattices with temporal  extent $N_\tau =6$  \cite{Cheng:2007jq}. These are the lattice results closest to the continuum limit.
Our results for the bulk viscosity in units of the entropy density
are shown in Fig.~\ref{fig:viscosity}. We note that the contribution
from the fermion sector of the trace anomaly is at least a factor 3 smaller than the gluonic contribution. Once lattice calculations with
physical quark masses are performed we expect this contribution to become
even less important. In Fig.~\ref{fig:viscosity}(right) we show results for $\zeta/s$ using  $500~{\rm MeV} \le \omega_0 \le 1500~{\rm MeV}$.

\subsection{Uncertainties of the method}

Our analysis relies on lattice results for the equation of state. Recent calculations with ${\cal O} (a^2)$ improved actions yield
quite consistent results \cite{Cheng:2007jq,Bernard:2006nj}, however
it should be noted that they are still not extrapolated to the continuum
limit and the light quark masses used in the calculation are about
a factor two larger than the physical values. 
Since the sum rule we use is exact (up to the terms quadratic in light quark masses), the main source 
of uncertainty is the ansatz for the spectral density. We have made the simplest possible assumption 
about its shape consistent with general physical requirements -- in particular, the spectral function must be linear for small frequencies. We also expect that the spectral density entering our sum rule 
should vanish above a certain frequency $\omega_0 \gg T$ at which the spectral density becomes perturbative and temperature--independent. 

We have estimated this frequency using lattice results on the temperature dependence of the running coupling \cite{Kaczmarek:2004gv,kz};
 $\omega_0$ is taken to be the inverse distance at which this running
coupling becomes approximately temperature independent and runs according
to the zero temperature $\beta$-functions. Choosing this scale involves
some uncertainty. Moreover, we have to admit
 that there is a sizable uncertainty in our ansatz. A different functional form 
of the spectral density would change the numerical value of extracted bulk viscosity. Fortunately, 
the temperature dependence of bulk viscosity is not sensitive to this  uncertainty, and so our main conclusion about the rapid growth of this quantity near the phase transition is robust. 
Moreover, we note that the spectral density extracted from the analyses of correlation functions 
on the lattice \cite{Karsch:2001uw,Meyer:2007dy} is quite similar to our ansatz.

\section{Universality}\label{unisec}

\subsection{The case of zero baryon density}

Let us discuss here a generic second order phase transition as
it might occur in the chiral limit of QCD at fixed strange quark mass
and most likely does occur in two-flavor QCD.
The critical behavior of thermodynamic quantities is in general 
controlled by two external parameters, the reduced temperature and the baryon chemical potential.

In the vicinity of the critical point the behavior of bulk
thermodynamic quantities is governed by thermal ($y_t$) and
magnetic $(y_h)$ critical exponents, which characterize the
scaling behavior of the singular part of the free energy density,
\begin{equation}
f(t,h)\equiv - \frac{T}{ V} \ln Z =b^{-1}f(b^{y_t}t, b^{y_h} h)~~.
\label{free}
\end{equation}
Here $b$ is an arbitrary scale factor. It is expected that the
chiral phase transition in QCD can be described by an effective,
three dimensional theory for the chiral order parameter, which in
the case of two-flavor QCD would amount to an $O(4)$ symmetric
spin model.

The scaling behavior of the specific heat 
is controlled by the critical exponents $\alpha = {(2y_t-1)/y_t}$.
Exponents relevant for a discussion of other susceptibilities as well
as the quark mass dependence of these quantities are
$\beta =(1-y_h)/y_t$, $\gamma = y_t/y_h$ and
$\delta=y_h/(1-y_h)$. Their numerical values for $Z(2)$ and
$O(4)$ symmetric spin models in three dimensions are given in
Table~\ref{table}.

\begin{table}[t]
\begin{center}
\vspace{0.3cm}
\begin{tabular}{|c|c|c|c|c|}
\hline
model & $\alpha$ & $\beta$ & $\gamma$ & $\delta$ \\
\hline
$O(4)$ & -0.21 & 0.38 & 1.47 & 4.82  \\
$Z(2)$ & 0.11 & 0.33 & 1.24 & 4.79 \\
\hline
\end{tabular}
\end{center}
\caption{
Critical exponents of 3-d $O(4)$ \cite{Engels} using $\beta$ and $\delta$
as input and $Z(2)$ \cite{Caselle} symmetric spin models.}\label{table}
\end{table}

Eq.~\eq{free} can be used to extract scaling laws for various
quantities, valid in the vicinity of the critical point. We are 
at present only interested in the scaling behavior of the specific
heat as function of temperature for vanishing external field 
\begin{equation}
c_v (t)  = c\ t^{-\alpha} + {\rm const}  \; ,
\label{cvscaling}
\end{equation}
which is obtained from Eq.~\eq{free} after taking two derivatives
with respect to $t$ and choosing $b=t^{-1/y_t}$; a constant $c$ can be both positive and negative.
In the $O(4)$ universality class the exponent $\alpha$ is negative.
The specific heat thus will not diverge but will only have a cusp.
From the relations given by Eq.~\eq{sum} we conclude that
similarly the bulk viscosity will not diverge but will have a maximum at 
$T_c$ and the velocity of sound will not vanish but will only attain a 
minimum.

\subsection{Chiral Critical Point}

The situation may be different at the chiral critical point, {\it i.e.}
the second order phase transition point that might exist in the QCD
phase diagram at non-vanishing chemical potential \cite{stephanov}. If it exists, this
critical point belongs to the universality class of the 3-d Ising model,
which has a positive specific heat exponent $\alpha$. The situation here,
however, is a bit more complicated as the energy-like and magnetization-like directions of the effective Ising model do not coincide with the
temperature and symmetry breaking (quark mass) directions of QCD, nor is
the latter controlled by the baryon chemical potential. Derivatives 
of the partition function with respect to temperature, which give the
specific heat, thus will usually be related to mixed derivatives with
respect to the energy-like and magnetization-like direction of the
effective Ising-Hamiltonian characterizing the universal behavior in the
vicinity of the chiral critical point. When approaching this critical 
point on a generic trajectory typical in a heavy ion collision, say for 
fixed $s/n_B$, the critical behavior
thus will not be controlled by $\alpha$ but by $\gamma/\beta\delta$ 
which will give the dominant singular behavior \cite{stephanov}.
This leads to an even more rapid divergence of the specific heat,
$c_v \sim h^{-\gamma/\beta\delta}$ with $\gamma/\beta\delta \simeq 0.8$
and where $h= A|T-T_c|/T_c + B |\mu_B - \mu_c|/\mu_c$ parametrizes the 
distance to the critical point. 

As should be clear from our discussion given in Section \ref{latsec} this also
implies that the bulk viscosity will diverge at the critical point and
the velocity of sound will vanish. Numerical evaluation of bulk viscosity close 
to chiral critical point would require lattice data at finite $\mu_B$. The sum rule in this case 
would also get modified due to the addition of terms containing derivatives in the 
chemical potential. We do not attempt such a numerical analysis in the present paper. 
However, the arguments for the divergence of bulk viscosity in the vicinity of chiral critical point 
which we presented above are sufficiently general.

\section{Summary and discussion}

Our results for the bulk viscosity obtained by combining low energy
theorems with lattice results for QCD with a physical strange quark mass
and almost physical light quark masses, indicate that the behavior is
qualitatively similar to the one observed previously for the case of $SU(3)$ gluodynamics -- 
the bulk viscosity rises dramatically in the vicinity of the phase transition. 
Therefore the increase of bulk viscosity near the phase transition is a sufficiently general phenomenon 
and is not specific only to the first-order phase transition in $SU(3)$ gluodynamics. 
 
We have argued that at zero baryon number density and in the limit of
vanishing light quark masses the bulk viscosity will develop universal
critical behavior which is related to that of the specific heat and will
belong  to the 3-dimensional, O(4) universality class if the transition
is second order in the chiral limit. The bulk viscosity in this case does not diverge at  $(T,\mu)=(T_c,\mu_c)$, but has a cusp.
 On the other hand, in the vicinity of the chiral critical point at finite baryon density, the critical behavior is in the $Z(2)$ universality class. This means that the bulk viscosity should {\it diverge} at $T=T_c$. 
In both cases the growth of bulk viscosity can be attributed to the excitation of a massless scalar mode 
responsible for long--distance correlations. The rather simple relation of the critical behavior of the
bulk viscosity to the (static) universality classes of QCD phase transitions
arises from the relation to bulk thermodynamic observables generated by the low energy theorems.
This is in contrast to the dynamic critical universality classes discussed
in the context of liquid-vapor transitions \cite{he3,Hohenberg}.

Large bulk viscosity signals strong coupling between the dilatational modes of the system and 
its internal degrees of freedom. 
We note that while for most physical substances the bulk viscosity is smaller or at most of the same order as the shear viscosity, this is not always true. For example, $^3$He in the vicinity of the critical liquid--vapor point exhibits the ratio of bulk-to-shear viscosities in excess of a \emph{million} \cite{he3}. 

What are the implications of our results? Since the growth of bulk viscosity signals an increase in entropy, our results imply that the expansion of hot QCD matter close to the phase transition is 
accompanied by the production of a large number of soft partons. These partons are 
produced so that the expanding quark-gluon system can hadronize -- the produced partons screen the color charges of the quarks and gluons originally present in the system. We thus come to the picture 
of ``soft statistical hadronization" -- ``soft" because the produced partons carry low momenta, and ``statistical" because the hadronization pattern is unlikely to depend upon the phase space distributions of ``pre-existing" partons -- the memory about these distributions is largely erased by the produced entropy. One may speculate that the association of inherent entropy with the hadronization process is analogous to the ``black hole hadronization" scenario, in which confinement is associated with an event horizon for colored partons \cite{Kharzeev:2005iz,Kharzeev:2006zm,Castorina:2007eb}.

In the vicinity of the chiral critical point, the divergence of bulk viscosity should  manifest itself in heavy ion collisions through the decrease of average transverse momentum of produced particles, accompanied by an increase in total multiplicity. This is due to both the increase in entropy associated with a large bulk viscosity, and the associated quenching of the transverse hydrodynamical  expansion of the system (``radial flow").

\vskip0.3cm
{\bf Acknowledgments.}
The work of F.K. and D.K. was supported by the U.S. Department of Energy under Contract No. DE-AC02-98CH10886. K.T. was supported in part by the U.S. Department of Energy under Grant No. DE-FG02-87ER40371; he would like to thank RIKEN, BNL, and the U.S. Department of Energy (Contract No. DE-AC02-98CH10886) for providing facilities essential for the completion of this work.



\begin{thebibliography}{70}

\bibitem{Kharzeev:2007wb}
  D.~Kharzeev and K.~Tuchin,
  arXiv:0705.4280 [hep-ph].

\bibitem{Boyd:1996bx}
  G.~Boyd, J.~Engels, F.~Karsch, E.~Laermann, C.~Legeland, M.~Lutgemeier and B.~Petersson,
  Nucl.\ Phys.\  B {\bf 469}, 419 (1996)
  [arXiv:hep-lat/9602007].

\bibitem{Meyer:2007dy}
  H.~B.~Meyer,
  arXiv:0710.3717 [hep-lat].

\bibitem{Nakamura:2007cm}
  S.~S.~A.~Nakamura,
  arXiv:0710.3625 [hep-lat].

\bibitem{Cheng:2007jq}
  M.~Cheng {\it et al.},
  arXiv:0710.0354 [hep-lat].


\bibitem{scale1}
  J.~R.~Ellis,
  Nucl.\ Phys.\  B {\bf 22}, 478 (1970);\\
  R.~J.~Crewther,
  Phys.\ Lett.\  B {\bf 33}, 305 (1970); \\
   M.~S.~Chanowitz and J.~R.~Ellis,
  Phys.\ Lett.\  B {\bf 40}, 397 (1972);\\
  J.~Schechter,
  Phys.\ Rev.\  D {\bf 21}, 3393 (1980).
\bibitem{scale2}
J.~C.~Collins, A.~Duncan and S.~D.~Joglekar,
  Phys.\ Rev.\  D {\bf 16}, 438 (1977);\\
  N.~K.~Nielsen,
  Nucl.\ Phys.\  B {\bf 120}, 212 (1977).

\bibitem{Novikov:1981xj}
  V.~A.~Novikov, M.~A.~Shifman, A.~I.~Vainshtein and V.~I.~Zakharov,
  Nucl.\ Phys.\  B {\bf 191}, 301 (1981).

\bibitem{Migdal:1982jp}
  A.~A.~Migdal and M.~A.~Shifman,
  Phys.\ Lett.\  B {\bf 114}, 445 (1982).
  
  \bibitem{KLT-Pom} 
  D.~Kharzeev, E.~Levin and K.~Tuchin,
  Phys.\ Lett.\ B {\bf 547}, 21 (2002)
  [arXiv:hep-ph/0204274];
  Phys.\ Rev.\ D {\bf 70}, 054005 (2004)
  [arXiv:hep-ph/0403152].
  
\bibitem{Bernard:1996cs}
  C.~W.~Bernard {\it et al.}  [MILC Collaboration],
  Phys.\ Rev.\  D {\bf 55}, 6861 (1997)
  [arXiv:hep-lat/9612025].

\bibitem{Aoki}
Y.~Aoki, Z.~Fodor, S.~D.~Katz and K.~K.~Szabo,
\emph{JHEP} {\bf 0601}, 089 (2006). 

\bibitem{Bernard:2006nj}
  C.~Bernard {\it et al.},
  Phys.\ Rev.\  D {\bf 75}, 094505 (2007)
  [arXiv:hep-lat/0611031].

\bibitem{Bringoltz:2005rr}
  B.~Bringoltz and M.~Teper,
  Phys.\ Lett.\  B {\bf 628}, 113 (2005)
  [arXiv:hep-lat/0506034].


\bibitem{Ellis:1998kj}
  P.~J.~Ellis, J.~I.~Kapusta and H.~B.~Tang,
  Phys.\ Lett.\  B {\bf 443}, 63 (1998)
  [arXiv:nucl-th/9807071].


\bibitem{Shushpanov:1998ce}
  I.~A.~Shushpanov, J.~I.~Kapusta and P.~J.~Ellis,
  Phys.\ Rev.\  C {\bf 59}, 2931 (1999)
  [arXiv:nucl-th/9901033].

\bibitem{Meyer:2007fc}
  H.~B.~Meyer,
  arXiv:0711.0738 [hep-lat].

\bibitem{Kaczmarek:2004gv}
  O.~Kaczmarek, F.~Karsch, F.~Zantow and P.~Petreczky,
  Phys.\ Rev.\  D {\bf 70}, 074505 (2004)
  [Erratum-ibid.\  D {\bf 72}, 059903 (2005)]
  [arXiv:hep-lat/0406036].

\bibitem{kz}
O.~Kaczmarek and F.~Zantow,
  Phys.\ Rev.\  D {\bf 71}, 114510 (2005)
  [arXiv:hep-lat/0503017].
  
\bibitem{Karsch:2001uw}
  F.~Karsch, E.~Laermann, P.~Petreczky, S.~Stickan and I.~Wetzorke,
  Phys.\ Lett.\  B {\bf 530}, 147 (2002)
  [arXiv:hep-lat/0110208]; P. Petreczky, {\it private communication}.

\bibitem{Engels}
J.~Engels, L.~Fromme and M.~Seniuch,
Nucl.\ Phys.\  B {\bf 675}, 533 (2003)
[arXiv:hep-lat/0307032].

\bibitem{Caselle}
M.~Caselle and M.~Hasenbusch,
J.\ Phys.\ A  {\bf 30}, 4963 (1997)
[arXiv:hep-lat/9701007].



  \bibitem{stephanov}
M. Stephanov, K. Rajagopal and E. Shuryak, Phys. Rev. Lett. {\bf 81}
(1998) 4816.

\bibitem{he3}
  A. Kogan and H. Meyer, J. Low Temp. Phys. {\bf 110} (1998) 899.
  
\bibitem{Kharzeev:2005iz}
  D.~Kharzeev and K.~Tuchin,
  Nucl.\ Phys.\  A {\bf 753}, 316 (2005)
  [arXiv:hep-ph/0501234].

  
\bibitem{Kharzeev:2006zm}
  D.~Kharzeev, E.~Levin and K.~Tuchin,
  Phys.\ Rev.\  C {\bf 75}, 044903 (2007)
  [arXiv:hep-ph/0602063].

\bibitem{Castorina:2007eb}
  P.~Castorina, D.~Kharzeev and H.~Satz,
  Eur.\ Phys.\ J.\  C {\bf 52}, 187 (2007)
  [arXiv:0704.1426 [hep-ph]].

\bibitem{Hohenberg}
P.C. Hohenberg and B.I. Halperin, Rev.Mod. Phys. {\bf 49}, 435 (1977).
\end{thebibliography}
\end{document}